\documentclass{ws-ijmpa}

\usepackage{graphics}
\usepackage{amssymb}
\usepackage{amsfonts}
\usepackage{amsmath}
\usepackage[T2A]{fontenc}

\usepackage{graphics}
\usepackage{epsfig}

\begin{document}

\markboth{}
{Thermodynamics of NJL-like models}

%
\catchline{}{}{}{}{}
%

\title{THERMODYNAMICS OF NJL-LIKE MODELS}

\author{FRIESEN A.V.}

\address{Bogoliubov Laboratory of Theoretical Physics, Joint Institute for 
Nuclear Research, 141980 Dubna, Russia\\
avfriesen@theor.jinr.ru}

\author{KALINOVSKY Yu.L.}

\address{Laboratory of Information Technologies, Joint Institute for Nuclear 
Research, 141980 Dubna, Russia \\
Higher Mathematics Department, University "Dubna", Dubna, Russia \\
kalinov@jinr.ru}

\author{TONEEV V.D.}

\address{Bogoliubov Laboratory of Theoretical Physics, Joint Institute for 
Nuclear Research, 141980 Dubna, Russia\\
toneev@theor.jinr.ru}

\maketitle

\begin{history}
\received{Day Month Year}
\revised{Day Month Year}
\end{history}

\begin{abstract}
The thermodynamic behavior of conventional Nambu-Jona-Lasinio and
Polyakov-loop-extended Nambu-Jona-Lasinio models is compared. 
Particular attention is paid to the phase diagram in the $T -\mu$
plane.

\end{abstract}

\ccode{PACS numbers:  11.30.Rd, 12.20.Ds, 14.40.Be}

\section{Introduction}

Models of the Nambu--Jona-Lasinio (NJL) type~\cite{nambu,volk} have a long 
history and have been extensively used to describe the dynamics of lightest 
hadrons and the thermodynamic properties of excited matter (see the review 
articles \cite{VW91,eqpi,HK94,buballa,volk1}). 
In the ``classical'' versions such schematic models incorporate the chiral 
symmetry of two-flavor QCD and its spontaneous breakdown at temperatures 
below the critical one, $T<T_c$.
They offer a simple and practical illustration of the basic mechanisms that 
drive the spontaneous breaking of chiral symmetry, a key feature of QCD in its 
low-temperature, low-density phase.
However, in spite of their widespread use, the NJL models suffer from the major
shortcoming that the reduction to global color symmetry prevents quark 
confinement.

In the Polyakov-loop-extended NJL (PNJL) model~\cite{MMO02}\cdash\cite{GMMR06},
the quarks are coupled simultaneously to the chiral condensate, to be an order 
parameter of the chiral symmetry breaking, and to a homogeneous gauge field
representing Polyakov loop dynamics, which serves as an order parameter for 
the transition from the low-temperature, symmetric, confined phase to the 
high-temperature, deconfined phase. 
The model has proven successful in reproducing lattice data on
QCD thermodynamics~\cite{ratti}.

In this paper we confront general properties of $\pi$ and $\sigma$ mesons
as well as thermodynamics at finite temperature $T$ and baryon chemical 
potential $\mu$ calculated within the two-flavor NJL model with those of the 
PNJL one.

\subsection{Nambu--Jona-Lasinio model}

To describe the coupling between quarks and the chiral condensate
in the scalar-pseudoscalar sectors, the two-flavor NJL
model~\cite{nambu,buballa,pot,klevansky} is used with the
following Lagrangian density
\begin{eqnarray}\label{njl}
\mathcal{L}_{\rm NJL}=\bar{q}\left(i\rlap/\partial-\hat{m}_0
\right) q+ G \left[\left(\bar{q}q\right)^2+\left(\bar{q}i\gamma_5
\vec{\tau} q \right)^2\right]~,
\end{eqnarray}
where $G$ is the coupling constant, $\vec{\tau}$ is the vector of Pauli 
matrices in flavor space, $\bar{q}$ and $q$ are the quark fields (color and 
flavor indices are suppressed),
$\hat{m}_0$ is the diagonal matrix of the current quarks masses,
$\hat{m}_0 = \mbox{diag} \, (m^0_u, m^0_d)$, $m^0_u = m^0_d=m_0$.

The grand canonical thermodynamic potential can be obtained from this 
Lagrangian in systematic approximations. 
In the mean-field approximation it has the form~\cite{pot}

\begin{eqnarray}
\Omega_{\rm NJL} = G\langle \bar q q \rangle ^2 + \Omega_q~,
\label{potnjl}
\end{eqnarray}
with
\begin{eqnarray}
\Omega_q = - 2N_cN_f\int \dfrac{d^3p}{(2\pi)^3}E_p 
- 2N_cN_fT\int \dfrac{d^3p}{(2\pi)^3}\left[\ln N^+(E_p) +\ln N^-(E_p)\right]~,
\end{eqnarray}
where $N^+(E_p)= 1 + e^{-\beta(E_p - \mu)}$ and 
$N^-(E_p) = 1 + e^{-\beta(E_p + \mu)}$ with 
$E_p = \sqrt{{\bf p}^2 + m^2}$ and the inverse temperature $\beta = 1/T$.

\subsection{Nambu--Jona-Lasinio model with Polyakov-loop}

The deconfinement in pure $SU(N_c)$ gauge theory can be
simulated by introducing an effective potential for a complex
Polyakov loop field. 
The PNJL Lagrangian~\cite{ratti,rosner}\cdash\cite{zhang} is
\begin{eqnarray}
\label{Lpnjl}
\mathcal{L}_{\rm PNJL}=\bar{q}\left(i\gamma_{\mu}D^{\mu}-\hat{m}_0
\right) q+ G \left[\left(\bar{q}q\right)^2+\left(\bar{q}i\gamma_5
\vec{\tau} q \right)^2\right]
-\mathcal{U}\left(\Phi[A],\bar\Phi[A];T\right)~.
\end{eqnarray}
Here, the notation is the same as in Eq.~(\ref{njl}).

The quark fields are coupled to the gauge field $A^\mu$ through the covariant 
derivative
$D^\mu = \partial^\mu -iA^\mu$.
The gauge field is $A^\mu = \delta_0^\mu A^0 = -i
\delta_4^\mu A_4$ (the Polyakov gauge).
The field $\Phi$ is determined by the trace of the Polyakov loop
$L(\vec{x})$~\cite{ratti}
\begin{eqnarray}
\Phi[A] = \dfrac{1}{N_c} \mbox{Tr}_c L(\vec{x})~,\nonumber
\end{eqnarray}
where $L(\vec{x}) = \mathcal{P} \exp \left[ \displaystyle
i \int_{0}^{\beta} d \tau A_4 (\vec{x}, \tau) \right]$.

The gauge sector of the Lagrangian density (\ref{Lpnjl}) is described by an
effective potential
$\mathcal{U}\left(\Phi[A],\bar\Phi[A];T\right)$
fitted to the lattice QCD simulation results in pure $SU(3)$ gauge theory at
finite $T$~\cite{ratti,rosner} with
\begin{eqnarray}\label{effpot}
\frac{\mathcal{U}\left(\Phi,\bar\Phi;T\right)}{T^4}
&=&-\frac{b_2\left(T\right)}{2}\bar\Phi \Phi-
\frac{b_3}{6}\left(\Phi^3+ {\bar\Phi}^3\right)+
\frac{b_4}{4}\left(\bar\Phi \Phi\right)^2, \\ \label{Ueff}
b_2\left(T\right)&=&a_0+a_1\left(\frac{T_0}{T}\right)+a_2\left(\frac{T_0}{T}
\right)^2+a_3\left(\frac{T_0}{T}\right)^3~.
\end{eqnarray}
The parameters of the effective potential (\ref{effpot}) and (\ref{Ueff}) are 
summarized in Table \ref{table1}.

\begin{table}[bh]
\tbl{  Parameters of the effective potential  $\mathcal{U}[A]$.}
{\begin{tabular}{@{}cccccc@{}} \toprule
$a_0$ & $a_1$ & $a_2$ & $a_3$ & $b_3$ & $b_4$ \\ \colrule
6.75 & -1.95 & 2.625 & -7.44 & 0.75 & 7.5 \\ \botrule
\end{tabular}\label{table1}}
\end{table}

The parameter $T_0$ in general depends on the number of active flavors and the
chemical potental \cite{Schaefer:2007pw}. 
In the present work we use $T_0 = 270$ MeV as in \cite{ratti}.

The  thermodynamic potential for the PNJL model in the mean-field 
approximation is given by the following equation~\cite{hansen}

\begin{eqnarray} 
\label{potpnjl}
\Omega (\Phi, \bar{\Phi}, m, T, \mu) &=&
\mathcal{U}\left(\Phi,\bar\Phi;T\right) + G \langle \bar{q}q \rangle ^2 
+\Omega_q~,
\end{eqnarray}
where (in analogy with (\ref{potnjl}))
\begin{eqnarray}
\Omega_q = -2 N_c N_f \int \dfrac{d^3p}{(2\pi)^3} E_p 
- 2N_f T \int \dfrac{d^3p}{(2\pi)^3} \left[ \ln N_\Phi^+(E_p)+
\ln N_\Phi^-(E_p) \right]~.
\end{eqnarray}
Here, $E_p=\sqrt{{\bf p}^2+m^2}$ is the quark energy, $E_p^\pm = E_p\mp \mu$, 
and
\begin{eqnarray}
&& N^+_\Phi(E_p) = \left[ 1+3\left( \Phi +\bar{\Phi} e^{-\beta
E_p^+}\right) e^{-\beta E_p^+} + e^{-3\beta E_p^+}
\right], \\
&& N^-_\Phi(E_p) = \left[ 1+3\left( \bar{\Phi} + {\Phi} e^{-\beta
E_p^-}\right) e^{-\beta E_p^-} + e^{-3\beta E_p^-} \right]~.
\end{eqnarray}

Since NJL-type models are nonrenormalizable it is necessary to introduce a 
regularization, e.g., by a cutoff $\Lambda$ in the momentum integration.
Following \cite{hansen}, we use in this study the three-dimensional
momentum cutoff for vacuum terms and extend this integration till
infinity for finite temperatures. 
A comprehensive study of the differences between the two regularization 
procedures (with and without cutoff on the quark momentum states at finite 
temperature) was performed in \cite{costa2}.

\section{Quarks and light mesons in NJL and PNJL models}

In the mean-field approximation, we can obtain the constituent
quark mass $m$ from the condition that the thermodynamical potential 
(Eqs.~(\ref{potnjl}) and (\ref{potpnjl}), resp.) shall have
a minimum  with respect to varying this parameter, 
$\partial \Omega/\partial m = 0$.
This condition  is equivalent to the gap equation~\cite{klevansky,hansen}
\begin{eqnarray} 
\label{gap1}
m = m_0 -2 G \ \langle \bar{q} q \rangle \label{masq}~,
\end{eqnarray}
where the quark condensate is defined as 
$\langle \bar{q} q \rangle = \partial \Omega/\partial m_0$.
For the mass gap equation of both models we get
\begin{eqnarray}
m = m_0 + 8 G N_c N_f \int_{\Lambda} \dfrac{d^3p}{(2\pi)^3}
\dfrac{m}{E_p} \left[ 1 - f^+ - f^- \right]~,
\end{eqnarray}
with
 \begin{eqnarray}
 f^+ &=& (1 + e^{\beta E_p^+})^{-1} \label{fermi},\\
 f^- &=& (1 + e^{\beta E_p^-})^{-1} \label{afermi}
 \end{eqnarray}
for the NJL model, and
\begin{eqnarray}
f^+ &=& \left[\left( \Phi +2\bar{\Phi} e^{-\beta E_p^+}\right) e^{-\beta E_p^+}
+ e^{-3\beta E_p^+}\right]/ N_\Phi^+(E_p)~, 
\label{fermimod}\\
f^- &=& \left[\left(\bar{\Phi}+2{\Phi} e^{-\beta E_p^-}\right) e^{-\beta E_p^-}
+ e^{-3\beta E_p^-} \right]/ N_\Phi^-(E_p) 
\label{afermimod}
\end{eqnarray}
for the PNJL model. 
Moreover, for PNJL calculations we should find the values of $\Phi$
and $\overline{\Phi}$ by minimizing $\Omega$ with respect $\Phi$
and $\overline{\Phi}$~\cite{hansen} at given $T$ and $\mu$. 
One should note, that if $\Phi \rightarrow 1$, the expressions 
Eqs.~(\ref{fermimod}),(\ref{afermimod}) reduce to the standard NJL 
model Eqs.~(\ref{fermi})  and (\ref{afermi}).

For a self-consistent description of the particle spectrum in the mean-field
approximation, the meson correlations have to be taken into consideration.
These correlations are related to the polarization operator of constituent 
fields. 
For scalar and pseudoscalar particles the polarization operators are 
represented by loop-integrals~\cite{eqpi,schulze,quack}.

\begin{eqnarray}
\Pi^{PP}_{ab} (P^2) &=& \int \frac{d^4p}{(2\pi)^4} \mbox{Tr}\,
\left[ i \gamma_5 \tau^a S(p+P) i \gamma_5 \tau^b S(p)
 \right], \label{Polpi} \\
\Pi^{SS}_{ab} (P^2) &=& \int \frac{d^4p}{(2\pi)^4} \mbox{Tr}\,
 \left[S(p+P) S(p) \right],
  \label{Polsig}
\end{eqnarray}
where the operation $\mbox{Tr}$ is taken over Dirac, flavor and color indices
of quark fields.

From point of view of the polarization operators, 
The pseudoscalar ($\pi$) and scalar ($\sigma$) meson masses can be defined by 
 the condition that for $P^2=M_\pi^2$ $(M_\sigma^2)$ the corresponding 
polarization operator $\Pi^{PP}(M_\pi^2)$ ($\Pi^{SS}(M_\sigma^2)$), leads to a
bound state pole in the corresponding meson correlation function \cite{hansen}.
For mesons at rest (${\mathbf P}=0$) in the medium, these conditions 
correspond to the equations
 \begin{eqnarray}
1 + 8 G N_c N_f \int \frac{d^3 p}{(2\pi)^3}
\frac{E_p}{M_\pi^2-4 E_p^2} \left( 1- f^+ - f^- \right) &=& 0, 
\label{masspi}\\
1 + 8 G N_c N_f \int \frac{d^3 p}{(2\pi)^3}\frac{1}{E_p} 
\frac{E_p^2-m^2}{M_\sigma^2-4 E_p^2} \left(1-f^+ -f^-\right) &=& 0~. 
\label{masssigma}
\end{eqnarray}

In order to solve Eqs.~(\ref{masq}), (\ref{masspi}) and (\ref{masssigma}), a 
set of model parameters has to be determined: the cutoff parameter $\Lambda$, 
the current quark mass $m_0$ (in chiral limit $m_0=0$) and the coupling
constant $G$. 
These parameters are fixed at $T = 0$ to reproduce physical quantities: the 
pion mass $M_\pi = 0.139$ GeV, the pion decay constant $F_\pi = 0.092$ GeV and 
the quark condensate $\langle \bar{q} q\rangle^{1/3}=-250$ MeV. 
The obtained parameters are shown in Table~\ref{table2}.

\begin{figure}[h]
\centerline{
\psfig{file = 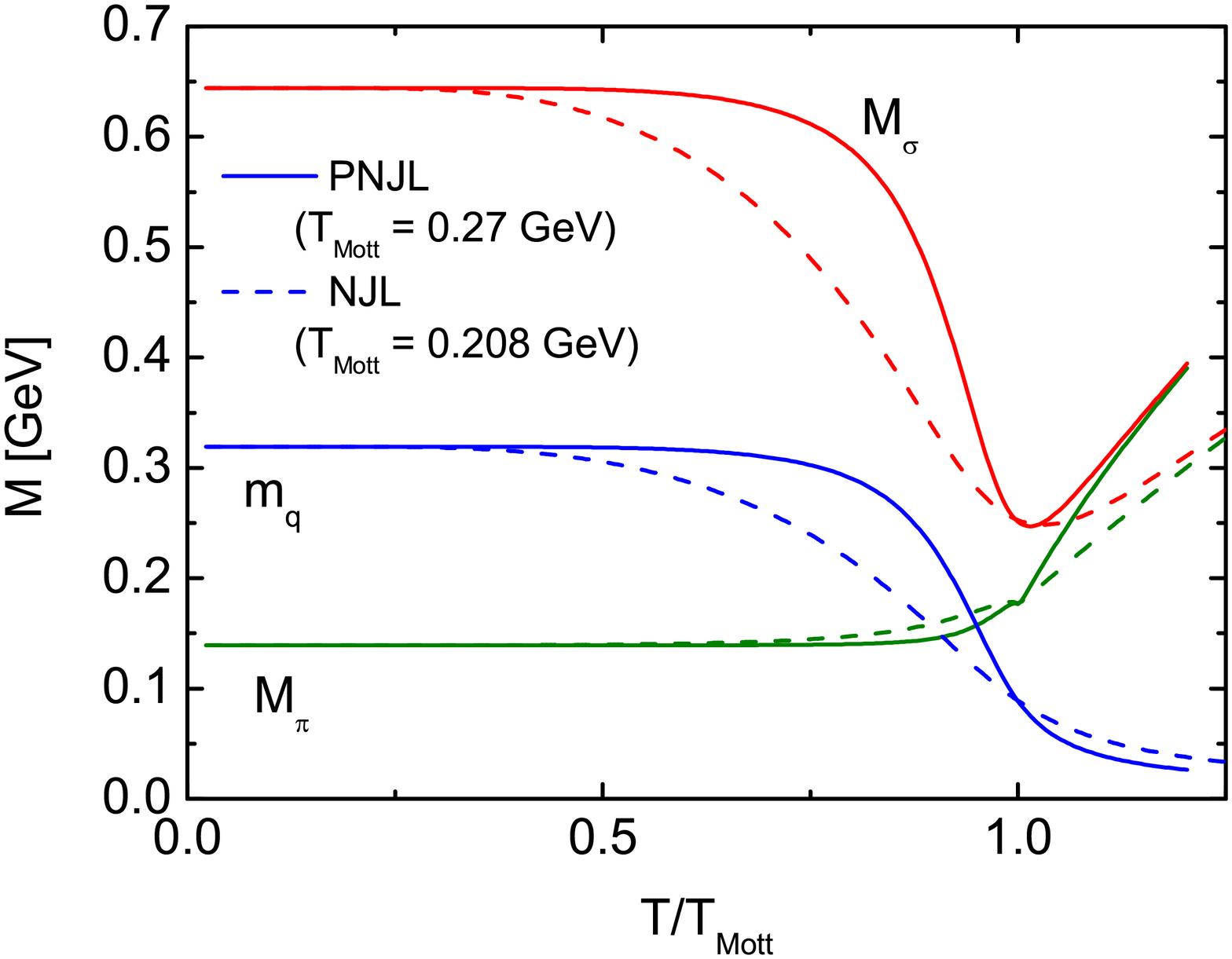, width = 7.5 cm}}
\caption{ Temperature dependence of the masses $m_q$, $M_\pi$ and $M_\sigma$
at $\mu$ = 0 GeV. Results for PNJL and NJL models are given
by the solid and dashed lines, respectively. }
 \label{masses}
\end{figure}

\begin{table}[ph]
\tbl{The set of model parameters reproducing observable quantities 
(in brackets) and the chiral condensate 
$\langle \bar{q} q\rangle^{1/3}=-250$ MeV.}
{\begin{tabular}{@{}ccccc@{}}\toprule
$m_0$ [MeV] & $\Lambda$ [GeV] & $G$ [GeV]$^{-2}$ & $F_\pi$ [GeV] & $M_\pi$ [GeV] \\ \colrule
5.5 & 0.639 & 5.227 & (0.092) & (0.139) \\ 
\botrule
\end{tabular}
 \label{table2}}
 \end{table}
Solutions of the gap-equation (\ref{masq}) and Eqs.~(\ref{masspi}), 
(\ref{masssigma}) at nonzero $T$ are presented in Fig.~\ref{masses}. 
The temperature 
is normalized to the Mott temperature, which is defined from the condition 
$M_\pi (T_{\rm Mott})= 2m_q(T_{\rm Mott})$. 
In the PNJL model $T_{\rm Mott} \simeq 0.27$ GeV and in the NJL model 
$T_{\rm Mott} \simeq 0.208$ GeV for our parameters.
The modification of the quasiparticle properties is clearly seen in this 
figure.  
Up to the Mott temperature $T_{\rm Mott}$, the $\sigma$ mass practically 
follows the behaviour of $2m_q(T)$ with a drop towards the pion mass 
signalling chiral symmetry restoration. 
At $T/T_{\rm Mott}>1$ the masses of chiral partners become equal to
each other, $M_\sigma \approx M_\pi$, and then both masses
increase with temperature.  
Below the Mott temperature, the pion mass remains practically constant. 
The transition region from the phase with broken chiral symmetry 
($m_q(T)\sim m_q(0)$) to chirally symetric phase ($m_q(T)\sim 0$) is much 
narrower in the PNJL case when compared to the NJL model.
For a recent discussion of this issue within the nonlocal PNJL model, see
\cite{Horvatic:2010md}.

\section{Thermodynamics of NJL and PNJL models }

The thermodynamics of particles is described in terms of the grand canonical 
ensemble which is related with the Hamiltonian $H$ as follows:
\begin{eqnarray}
 \label{can}
e^{-\beta V \Omega} = \mbox{Tr}\,\, e^{-\beta (H-\mu N)},
\end{eqnarray}
where $N$ is the particle number operator and the operator
$\mbox{Tr}$ is taken over momenta as well as color, flavor and Dirac indices. 
If $\Omega$ is known, the basic thermodynamic quantities - the pressure, the 
energy and entropy densities, the density of quarks number and heat
conductivity - can be defined as follows:
\begin{eqnarray}
p &=& -\frac{\Omega}{V}, \\
s &=& -\left(\frac{\partial \Omega}{\partial T}\right)_\mu, \\
\varepsilon &=& -p + Ts +\mu \, n, \\
n &=& -\left(\frac{\partial \Omega}{\partial \mu}\right)_T, \\
c &=& \frac{T}{V}\left(\frac{\partial s}{\partial T}\right)_\mu ~.
\end{eqnarray}
The thermodynamic potential in equilibrium corresponds to a global minimum 
with respect to variations of the order parameter(s)
\begin{eqnarray}
\frac{\partial \Omega (T,\mu,m) }{\partial m} = 0, \,\,\,
\frac{\partial^2 \Omega(T,\mu,m)}{\partial m^2} \geqslant 0.
\end{eqnarray}

All these relations describe thermodynamics of the system.
For the considered models the thermodynamic potentials are
defined from Eqs.~(\ref{potnjl})  and (\ref{potpnjl}). 
From these equations we can read off the vacuum part

\begin{eqnarray}
\label{omvac}
\Omega_{vac} = \frac{(m -m_{0})^2}{4G} - 2N_c N_f\int
\frac{d^3p}{(2\pi)^3}E_p.
\end{eqnarray}
This quantity does not vanish at $T\rightarrow 0$ and $\mu
\rightarrow 0$. 
Therefore, in order to obtain the physical thermodynamical 
potential which corresponds to vanishing pressure and energy density
at $(T,\mu)=(0,0)$, one has to renormalize the thermodynamic
potential by subtracting its vacuum expression (\ref{omvac}).
This corresponds to the following defintion of the physical pressure
\begin{eqnarray}
\frac{p}{T^4} = \frac{p(T, \mu, m) - p(0, 0, m)}{T^4}.
\end{eqnarray}

\begin{figure}[thb]
\centerline{
\psfig{file = 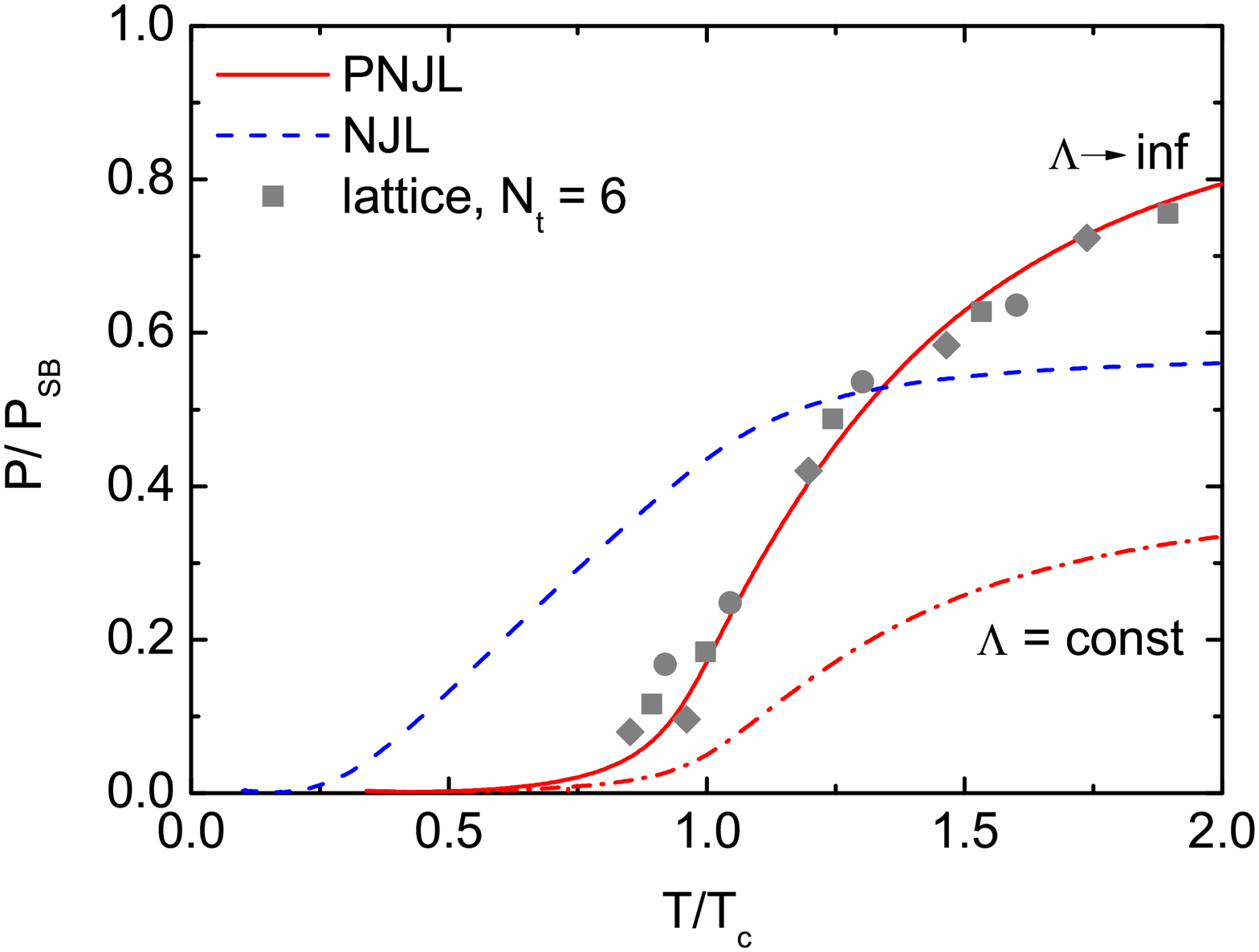, width=6.3cm}
\psfig{file = 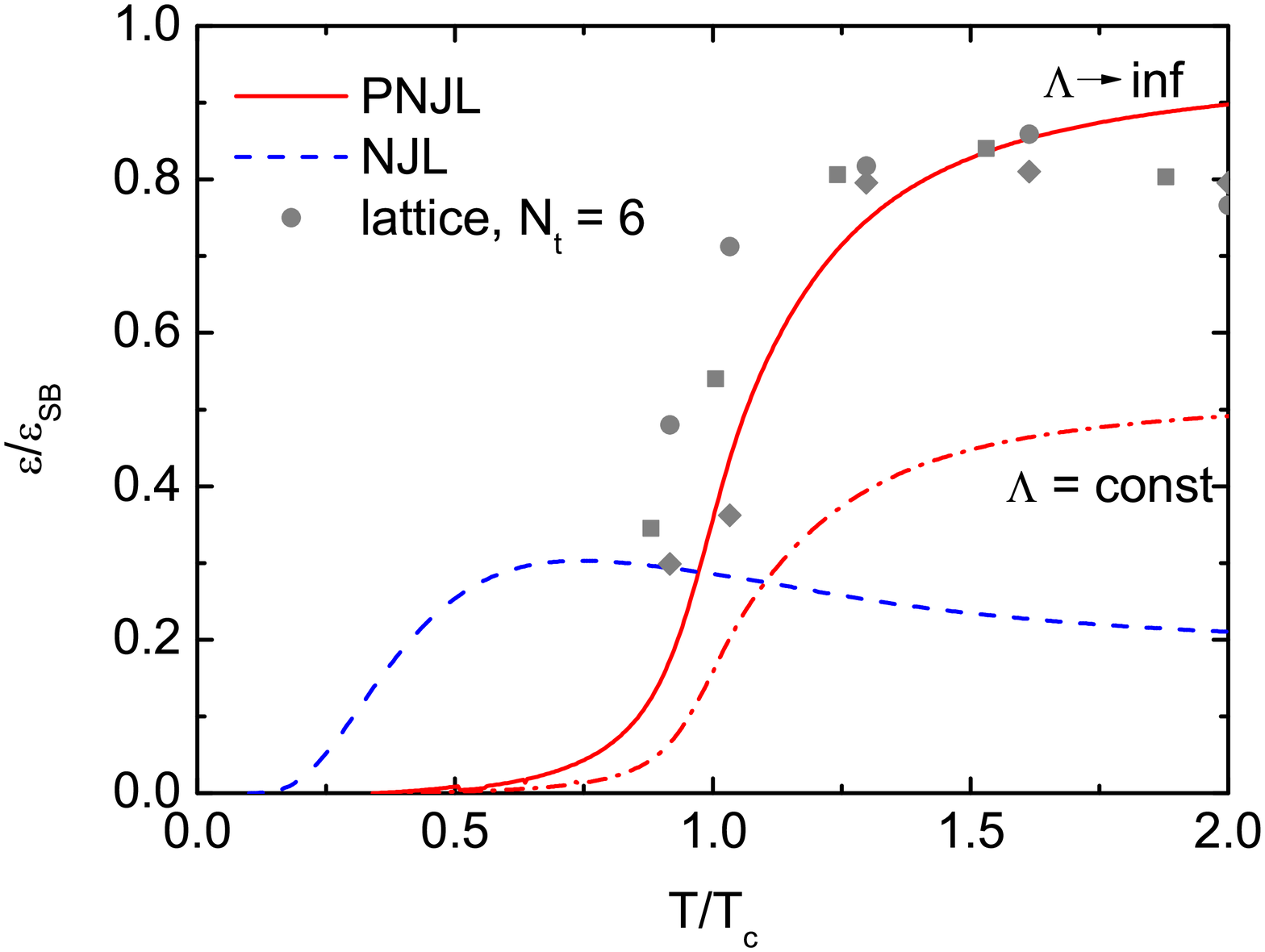, width=6.3cm} }
\caption{The temperature dependence of the reduced pressure
and energy density within the PNJL model for $\mu=0$
in two schemes of regularization $\Lambda =$0.639 and
$\Lambda\rightarrow\infty$. Dotted lines are
appropriate results for the NJL model.
Lattice data points for $N_f=2$ at $\mu=$0 are
from Ref.~[27].
Circles, squares and diamonds correspond to calculations at $N_t=$6
with the mass ratio of the pseudoscalar to vector meson
$m_{PS}/m_{V}=$0.65, 0.70 and 0.75, respectively.
}
\label{therm}
\end{figure}

Within the PNJL model with $\Lambda\rightarrow\infty$ the reduced
pressure and energy density exhibit reasonable behavior
consistent with the recent lattice QCD results for the vanishing
chemical potential~\cite{AKh01} (see Fig.~\ref{therm})  keeping in mind
that the $m_{PS}/m_{V}$ ratio in lattice calculations is still far from
that for physical masses $m_{PS}/m_{V}\sim 0.2$. One should note that
both models have the cutoff parameter $\Lambda$~\cite{hansen}. 
But the integrals containing the logarithm in Eq.~(\ref{potnjl}) and 
Eq.~(\ref{potpnjl}) are both convergent~\cite{ratti,sasaki}. 
In our work we calculated these integrals with $\Lambda\rightarrow\infty$. 
It leads to the flattening of pressure at high temperature. 
However, most integrals for the PNJL model are convergent too. 
This was the reason to consider the thermodynamic functions for 
$\Lambda \rightarrow \infty$.
It was supposed that with increasing temperature the pressure has to reach 
the Stefan-Boltzmann limit~\cite{costa2}, which in the PNJL model is defined as
\begin{eqnarray}
\frac{p_{SB}}{T^4} = (N_c^2 - 1)\frac{\pi^2}{45} + N_cN_f\frac{7\pi^2}{180}
\simeq 4.053.
\end{eqnarray}

If the regularization $\Lambda =$ 0.639 is used, the $T$-behaviour of the 
thermodynamic quantities considered is roughly the same while their absolute 
values are noticeably lower, being far from the Stefan-Boltzmann limit. 
In the NJL model both, $p/T^4$ and $\varepsilon/T^4$ are not only 
underestimated due to the missing gluon contribution, but also essentially 
shifted toward lower temperatures because of the lack of a confining mechanism
for the dynamical quark degrees of freedom.

\begin{figure}[thh]
\centerline{
\psfig{file = 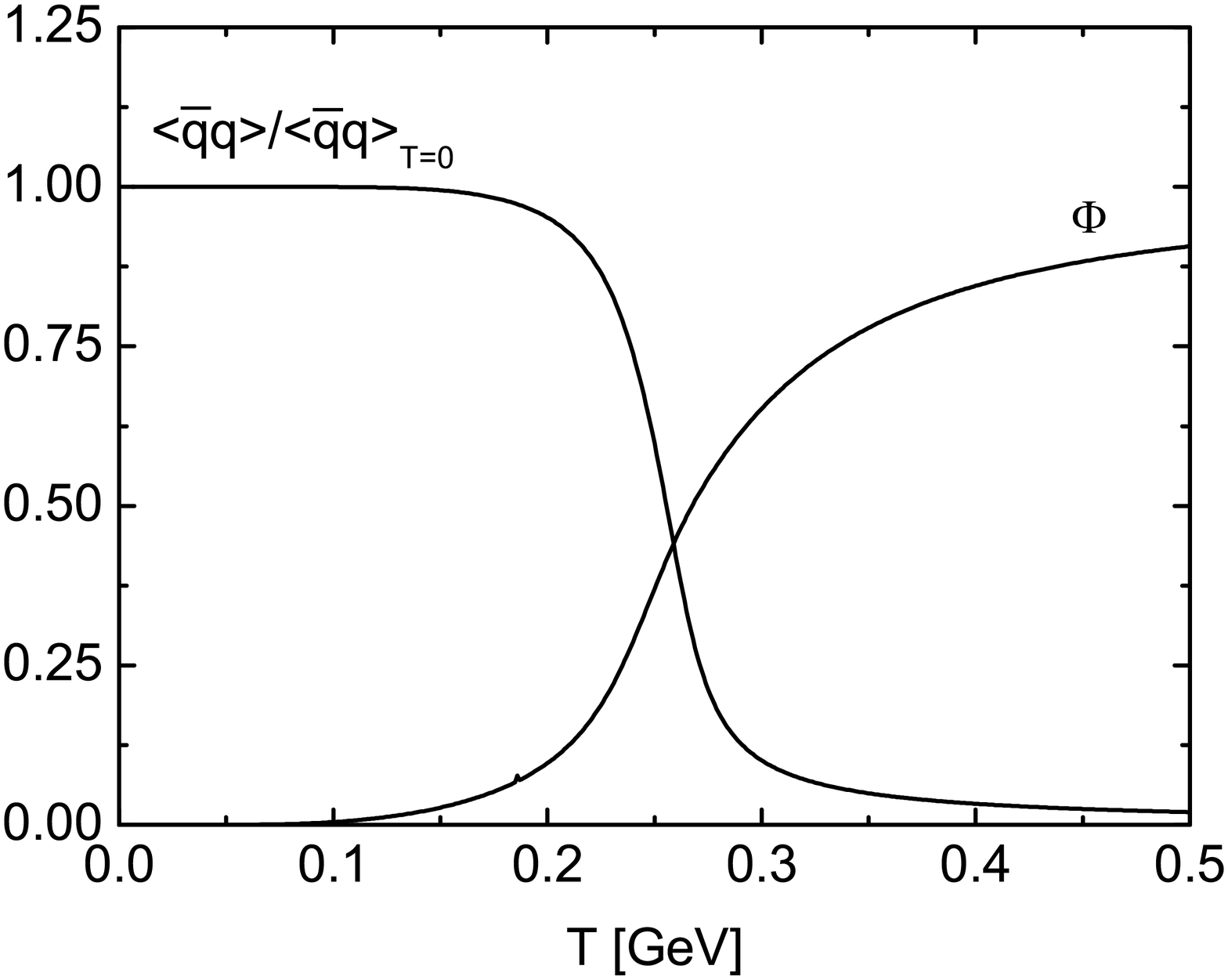, width = 6.5cm}}
\caption{ Temperature dependence of the chiral condensate and Polyakov loop
potential at $\mu$ = 0 GeV within the PNJL model }
 \label{order}
\end{figure}

Within NJL-like models there are several characteristic temperatures. 
The parameter $T_0$ entering the effective potential (\ref{Ueff}) of the PNJL 
model has been noted above. 
Three other scales are the pseudo-critical temperature for chiral crossover
$T_\chi$ defined by the maximum of 
$\partial \langle q\overline{q} \rangle /\partial T$, the pseudo-critical
temperature for the deconfinement crossover $T_p$ which can be found from the
maximum of $\partial \overline{\Phi}/\partial T$, and $T_c$
defined for PNJL model as the average of two transition
temperatures $T_\chi$ and $T_p$~\cite{ratti,rosner}. 
The temperature dependence of the order parameters for the chiral 
($\langle q\bar q \rangle $) and deconfinement ($\Phi$) phase transitions, are 
shown in Fig.~(\ref{order}).
The chiral condensate decreases and the Polyakov loop potential increases
with $T$, demonstrating closeness of the pseudo-critical temperatures $T_\chi$ 
and $T_p$ at $\mu=0$ (see also Table~\ref{table3}).
For $\pi$-mesons, the Mott temperature $T_{\rm Mott}$ is provided by
the condition $M_\pi(T_{\rm Mott}) = 2m_q(T_{\rm Mott})$ and similarly 
the $\sigma$ meson dissociation temperature $T_d^\sigma$ is given by the 
equation $M_\sigma(T_d^\sigma) = 2M_\pi(T_d^\sigma)$~\cite{quack,fu}. 
All these quantities obtained at $\mu=0$ are presented in Table \ref{table3}.

\begin{table}[thb]
\tbl{Characteristic temperatures in NJL and PNJL models.}
{\begin{tabular}{@{}ccccccc@{}} \toprule
 & $T_0$ & $T_\chi$ &$T_p $&$ T_c $&$T_{\rm Mott}$ & $T^\sigma_d$ \\\colrule
 NJL & \textendash& 0.192 & \textendash & 0.192 & 0.207 & 0.185\\
 PNJL & 0.27 & 0.249 & 0.258 & 0.253 & 0.27 & 0.259\\ \botrule
\end{tabular}
\label{table3}}
\end{table}

\begin{figure}[thb]
\centerline{
\psfig{file = 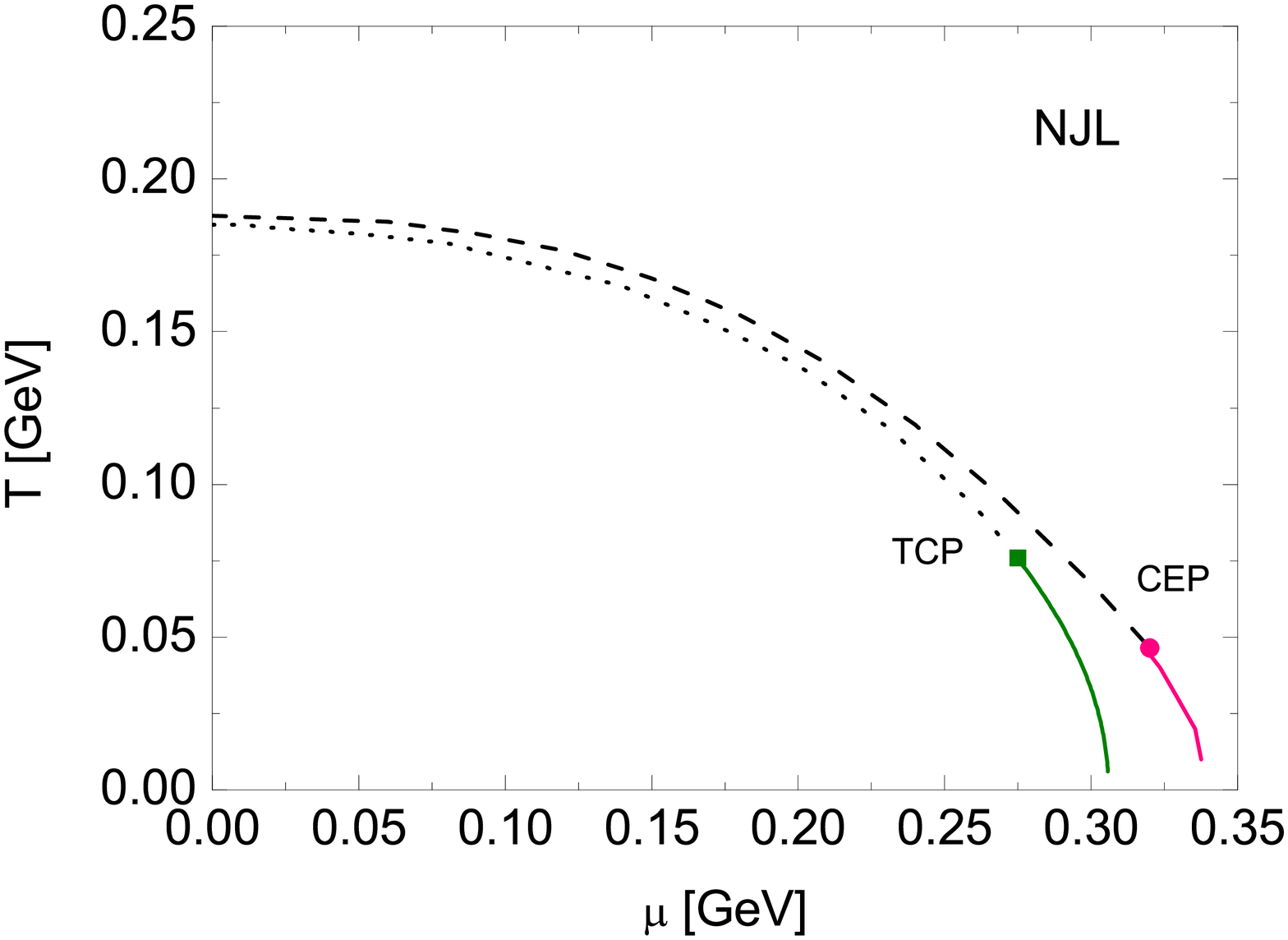,width = 6.5 cm}
\psfig{file = 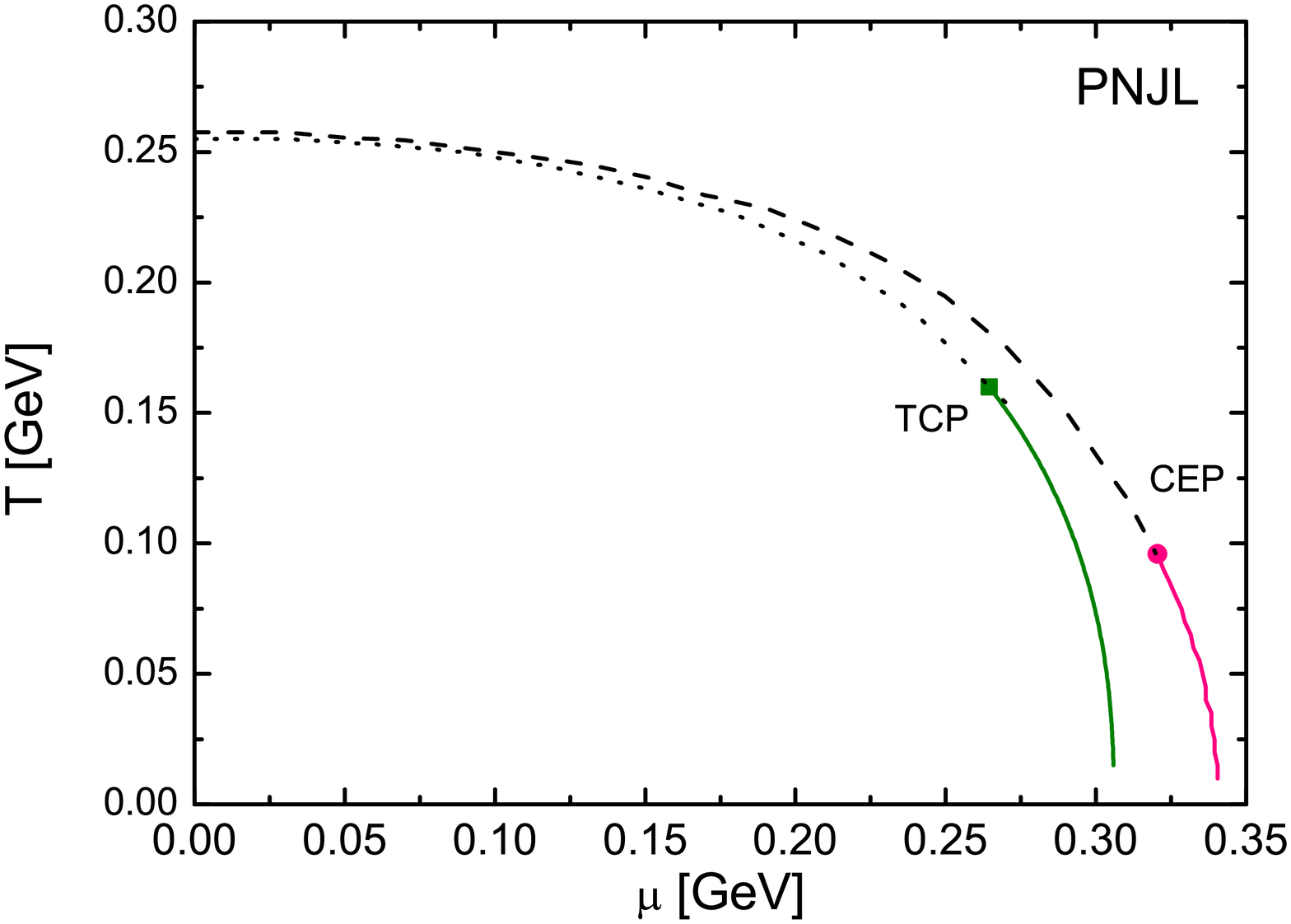, width = 6.5 cm}}
\caption{Phase diagrams of NJL (left panel) and PNJL (right panel)
models. Solid lines correspond to the first order phase
transition, dashed lines are crossover and dotted lines are the
boundary of the second order phase transition.}
 \label{phasediag}
\end{figure}

Extending our study of the (pseudo-)critical temperatures to nonzero chemical 
potential $\mu$, we obtain phase diagrams in the $T-\mu$ plane shown in 
Fig.~\ref{phasediag}.
The chiral transition line, determined by~\cite{eqpi}
$$\left[ \frac{1}{4G}+\frac{\partial \Omega_q}{\partial m^2}\right]_{m=0}=0$$
\noindent in both NJL and PNJL models, is a monotonously
decreasing function of the chemical potential. 
In the limiting chirally symmetric case corresponding to $m=0$ the $T_\chi$ at
large $\mu$ is higher than those for finite mass but both
temperatures coincide when $\mu \to$0 for both the NJL and PNJL models. 
For the case $\mu\ne$0 both models show the critical end point at the 
temperature $T_{CEP}$ below which the chiral phase
transition is of first order. 
At this point ($T_{CEP},\mu_{CEP}$) the phase transition changes from first
order
to crossover~\cite{costa2,kashiwa,fukushima}. 
For the chirally symmetric case in the PNJL model the first order phase 
transition ends at a tricritical point above which, for $T>T_{TCP}$, the 
chiral transition is of the second order. 
In the NJL model~\cite{SMMR01} the topology of the phase diagram is the same as
in the PNJL case, only $T_{CEP}$ and $T_{TCP}$ are situated at higher 
temperatures.
In accordance with other calculations, the temperature of
the tricritical point is above that of the critical end point.
Within the PNJL model the positions of critical points are
$(T_{CEP},\mu_{CP})=$(95,320) and $(T_{TCP},\mu_{TCP})=$(160,265)
MeV. 
These numbers are quite close to those in \cite{costa2} for the set with
similar parameter values (set B, the case I). 
One should emphasize that critical properties of observables are significantly 
influenced by the chosen parameter set and regularization procedure as was
demonstrated in \cite{costa2}.

\section{Conclusion}

We have compared the thermodynamics of NJL and PNJL models. 
In agreement with previous results, it is shown that the inclusion of coupling 
between chiral symmetry and deconfinement essentially improves the
description of thermodynamic bulk properties of the medium. 
The models qualitatively reproduce both, $\pi$ and $\sigma$ meson properties 
in hot, dense quark matter and the rich and complicated phase structure of this
medium. 
Effects of the Polyakov loop move the CEP to higher $T$ and lower $\mu$
than in the NJL case~\cite{rosner}. 
The position of the calculated CEP in the $T-\mu$ plane is still far from the 
predictions of lattice QCD and empirical analysis. 
The further elaboration of the presented models may include color 
superconducting phases and nonlocality of the interaction 
\cite{GomezDumm:2005hy} as well as effects beyond the meanfield 
\cite{Blaschke:2007np}.

\section*{Acknowledgments}

We are grateful to D. Blaschke, P. Costa and V. V. Skokov for useful comments. 
V.T. acknowledges financial support from the Helmholtz International Center
(HIC) for FAIR within the LOEWE program.
The work of Yu. K. was supported by RFFI grant No. 09-01-00770a.

\end{document}